\documentclass{article}
\usepackage{spconf,amsmath,graphicx}
\usepackage{siunitx}
\usepackage{float}
\usepackage{stfloats}

\newcommand{\crnotice}{\copyright 2024 IEEE. Personal use of this material is permitted. Permission from IEEE must be obtained for all other uses, in any current or future media, including reprinting/republishing this material for advertising or promotional purposes, creating new collective works, for resale or redistribution to servers or lists, or reuse of any copyrighted component of this work in other works.}
\newcommand{\pccy}{{\begin{tikzpicture} \node[text width=0.8\paperwidth, align=justify, anchor=south] {\crnotice}; \end{tikzpicture}}}
\usepackage{background}

\backgroundsetup{contents=\pccy, scale=1, angle=0, position={0.385\paperwidth, 15mm}, color=black, anchor=below}
\title{Detecting gamma-band responses to the speech envelope for the ICASSP 2024 Auditory EEG Decoding Signal Processing Grand Challenge}
\name{{Mike Thornton\textsuperscript{1}, Jonas Auernheimer\textsuperscript{2}, Constantin Jehn\textsuperscript{2}, Danilo Mandic\textsuperscript{1}, Tobias Reichenbach\textsuperscript{2}}\thanks{Mike Thornton is supported by UK Research and Innovation. [UKRI Centre for Doctoral Training in AI for Healthcare grant number EP/S023283/1].
The project was supported by the German Federal Ministry of Education and Research (SEMECO, 03ZU1210XX).}}
\address{\textsuperscript{1}Imperial College London, \textsuperscript{2}Friedrich-Alexander-Universit{\"a}t Erlangen-N{\"u}rnberg}
\begin{document}
\ninept
\maketitle

\begin{abstract}
%
%

The 2024 ICASSP Auditory EEG Signal Processing Grand Challenge concerns the decoding of electroencephalography (EEG) measurements taken from participants who listened to speech material. This work details our solution to the match-mismatch sub-task: given a short temporal segment of EEG recordings and several candidate speech segments, the task is to classify which of the speech segments was time-aligned with the EEG signals. We show that high-frequency gamma-band responses to the speech envelope can be detected with a high accuracy. By jointly assessing gamma-band responses and low-frequency envelope tracking, we develop a match-mismatch decoder which placed first in this task.

\end{abstract}
\section{Introduction}
\label{sec:intro}
The ICASSP Auditory EEG Signal Processing Grand Challenge (SPGC) focuses on the decoding of electroencephalography (EEG) measurements from participants who listened to speech. The 2023 edition of the SPGC included a binary classification problem called the match-mismatch task: given a short segment of EEG signals (three seconds in duration), and two candidate speech segments (also of three seconds duration), the task is to classify which of the speech segments is time aligned, or `matched', with the EEG segment~\cite{jalilpour2022icassp}. For this year's edition, the segment length was increased to five seconds. Additionally, the number of mismatched segments was increased from one to four, thereby increasing the task difficulty.

In the 2023 edition, we achieved the best score in the match-mismatch task through combining two decoders which exploited different EEG responses to speech: slow neural tracking of the speech envelope, and high-frequency speech-related frequency-following responses~\cite{Thorn2023a}. Both decoders shared a common deep learning architecture based on that of Accou~\textit{et al.}~\cite{Accou2021}; full details of these decoders are given in Thornton~\textit{et al.}~\cite{Thorn2023b}. By training multiple instances of the decoders with different weight initialisations and averaging their outputs, a further improvement in decoding accuracy was achieved. Another team showed that it was beneficial to select the mismatched speech segments randomly whilst training the decoders, rather than following the typical method of selecting mismatched segments which occurred at a fixed delay relative to the offset of the matched segments~\cite{Cui2023}.

\begin{figure}
    \centering
    \includegraphics[trim={0 20pt 0 15pt}, width=0.45\textwidth]{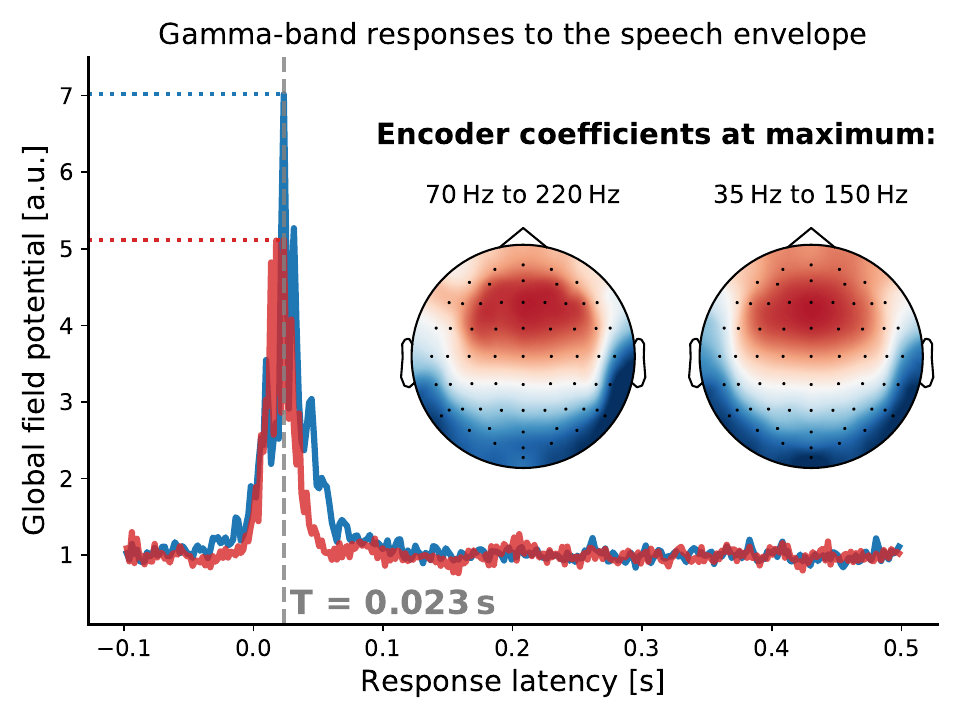}

    \caption{High-gamma (\SIrange[range-phrase=--]{70}{220}{Hz}, red) and lower-gamma (\SIrange[range-phrase=--]{35}{150}{Hz}, blue) envelope-related EEG temporal response functions. The high-gamma range captures the typical range of the fundamental frequency of speech. The global field potentials (GFPs) shown in the figure have been normalised so that the background activity has a value of 1. Therefore, the peak GFP values represent the SNRs of the two responses.}
    \label{fig:FFRs}
\end{figure}

For this year's edition of the SPGC, we adapted our previous approach by combining a low-frequency envelope-tracking decoder with a decoder which detects high-frequency (gamma-band) EEG responses to speech. Whereas we previously attempted to relate the EEG signals in the high-gamma range (\SIrange{70}{220}{Hz}) to the high-frequency envelope-modulations feature, we show here that the speech envelope itself can be related to EEG signals in the broad gamma range. Figure~\ref{fig:FFRs} shows that these gamma-band responses exhibit strong signal-to-noise ratios (SNRs) when frequencies as low as $\SI{35}{Hz}$ are considered.

\begin{table*}[t]
\centering
   \begin{tabular}{ |l|c|c|c| }
 \hline
  & \multicolumn{3}{c|}{\qty{5}{s} segment length} \\
 \hline
 Decoder & Absolute decoding & One imposter & Four imposters \\
 \hline
 Gamma-band & $69.52\pm 3.32$ & $77.43\pm 4.35$ & $53.51\pm 6.78$\\
 LF Envelope & $76.71\pm 2.79$ & $85.57\pm 3.20$ & $66.79\pm 6.01$\\
 Composite & $80.73\pm 2.75$ & $89.95\pm 2.74$ & $76.18\pm 5.37$\\
\hline
\end{tabular}
 \caption{The 95\% confidence intervals of the mean decoding accuracies of our decoders for 14 heldout participants from the training dataset.}
  \label{table:decoder-overview}
\end{table*}

\section{Materials and methods}
\label{sec:methods}

\textit{2.1. Dataset.} The SPGC organisers provided a dataset of EEG measurements taken from 85 participants who listened to speech material~\cite{K3VSND_2023}. Data from 71~participants were used to train the decoders, while the remaining 14~participants served for evaluating the decoders prior to submitting our solutions.

The pre-processing procedure followed that of Thornton~\textit{et al.}~\cite{Thorn2023b}. Here, due to the short segment length, all filtering operations were performed with low-order IIR filters to reduce edge artefacts. In brief, the low-frequency envelope decoder employed broadband EEG signals and speech envelopes which were resampled to a sampling rate of \SI{64}{Hz}. For the gamma-band envelope decoder, the EEG signals were filtered between \SIrange{35}{150}{Hz} and resampled to \SI{512}{Hz}; the broadband speech envelopes were also resampled to this rate.

\noindent\textit{2.2. Low-frequency (LF) envelope decoder. } Given an (EEG segment, envelope segment) pair with a sampling rate of \SI{64}{Hz}, this decoder produced a logit which describes how similar the two segments are. The decoder was trained in a binary classification setting using the Adam optimiser and the cross-entropy loss function. We used early stopping and a learning rate scheduler. The finding of Cui~\textit{et al.}~\cite{Cui2023} that it is beneficial to select mismatched segments at random during training was replicated.

\noindent\textit{2.3. Gamma-band envelope decoder. } The gamma-band envelope decoder shared the same architecture and hyperparameters as the LF envelope decoder, and was trained in the same way. However, it used the signals which were sampled at \SI{512}{Hz}.

\begin{figure}[h!]
    \centering
    \includegraphics[trim={0 45pt 0 20pt}, width=0.45\textwidth]{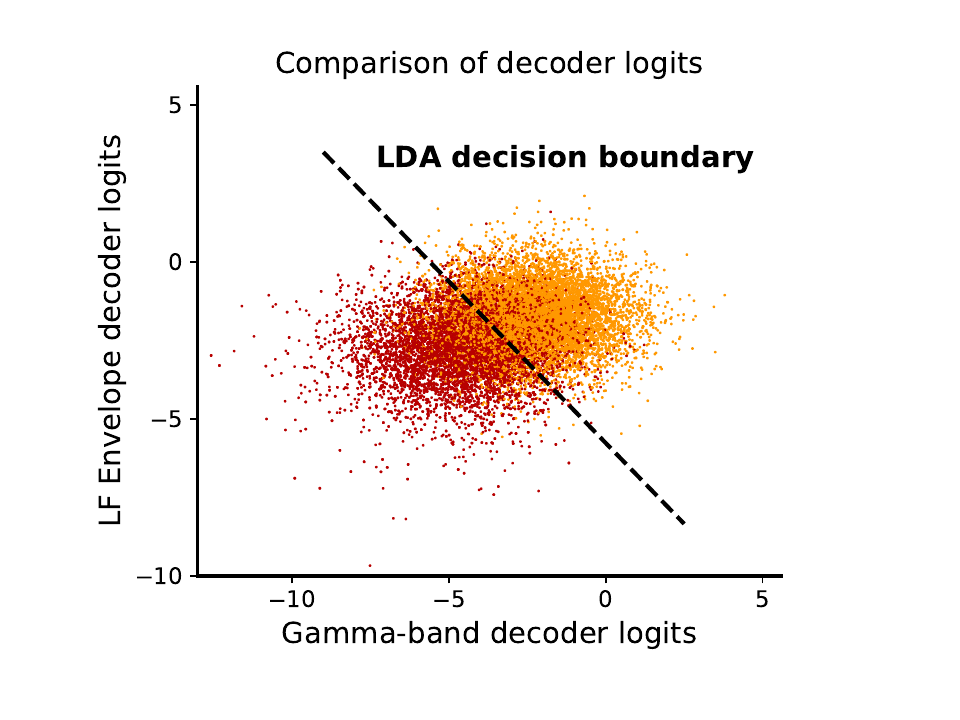}

    \caption{Comparison between the output logits of the low-frequency (LF) and gamma-band envelope decoders, as well as the decision boundary of the composite decoder. Orange dots represent matched segments, whereas red dots represent mismatched segments.}
    \label{fig:logits}
\end{figure}

\noindent\textit{2.4. Composite decoder. } The composite decoder was formed by combining the output logits of the LF envelope decoder and the gamma-band decoder via a linear classifier (linear discriminant analysis,  Figure~\ref{fig:logits}). This classifier produces an estimate of the probability that a particular (EEG segment, stimulus segment) pair is matched. When one pair of segments is known to be matched, and other pairings are imposters, the matched pair is estimated as that which produces the largest predicted probability.

\section{Results}
\label{sec:results}

Decoder training (including hyperparameter tuning) was performed using data from 71 of the participants in the training dataset. The remaining 14 participants were held-out for evaluation. Table~\ref{table:decoder-overview} shows the performances of the three decoders for three related tasks. In \textit{absolute decoding}, the decoders are required to classify whether a single (EEG segment, speech segment) pair is matched or not. In the other two tasks, a single EEG segment and several speech segments are provided, and the task is to classify which one of the speech segments is matched with the EEG.

Table~\ref{table:leaderboard} shows how our submitted solutions scored in the challenge. Our first submission achieved an accuracy of 59.97\%, placing our team second by a narrow margin. To improve our score, we re-used the decoder averaging technique which we employed last year. By separately averaging the output logits of 16 instances of each of the envelope and gamma-band decoders, our second solution achieved a considerably higher score of 62.80\%, ultimately placing our team first. 
The composite decoder achieved a much lower score than was expected based on our evaluation which used the 14 heldout participants from the training dataset, possibly suggesting some distributional shift between the two datasets; further investigations are required to understand the cause of this.

\begin{table}[H]
\centering
   \begin{tabular}{ |l|c|c| }
 \hline
 Decoder& Heldout accuracy & Leaderboard score\\
 \hline
 Single-instance & $76.18\pm 5.37$ & $59.97\pm 6.97$\\
 
 Multiple instances & $78.43\pm 5.21$ & $62.80\pm 6.88$\\
 \hline

\end{tabular}
\caption{The average accuracies (mean $\pm$ 95\% margin of error) of our submitted solutions. Our first submission made use of single-instance decoders. For our second submission, we averaged the output logits of 16 decoder instances before applying the LDA classifier.}
 \label{table:leaderboard}
\end{table}

\section{Conclusions}
\label{sec:discussion}

We have demonstrated gamma-band EEG responses to the speech envelope. Composite match-mismatch decoders, which monitor both low-frequency and gamma-band responses, have outperformed individual decoders based on either type of response. Responses to the low-gamma components of the speech envelope were particularly strong, and by incorporating these we have improved the classification accuracy of the gamma-band decoder. Finally, by averaging the output logits of multiple trained decoder instances, a considerable performance improvement has been observed.

\bibliographystyle{IEEEbib}
\bibliography{bibliography}

\end{document}